\documentclass[12pt,preprint]{aastex}

\def \etal         {{et~al. }}
\def  \h2        {\hbox{H$_2$}}

\def \kms          {\hbox{km$\,$s$^{-1}$}}
\def\approxlt{\lower.2em\hbox{$\buildrel < \over \sim$}}
\def\approxgt{\lower.2em\hbox{$\buildrel > \over \sim$}}
\def \lco          {\hbox{$L^{\prime}_{\rm CO}$}}
\def \lhcn          {\hbox{$L^{\prime}_{\rm HCN}$}}

\def \lir          {\hbox{$L_{\rm IR}$}}
\def \lfir          {\hbox{$L_{\rm FIR}$}}
\def \ls           {\hbox{L$_{\odot}$}}

\begin{document}
 
\title {HCN Observations of Dense Star-Forming Gas 
in High Redshift Galaxies}

\author {Yu Gao$^{1, 2
}$, \ 
Chris L. Carilli$^2$, \ Philip M. Solomon$^3$, and Paul A. Vanden Bout$^4$}
 
\altaffiltext{1}{Purple Mountain Observatory, Chinese Academy of Sciences 
(CAS), 2 West Beijing Road, Nanjing 210008, P.R. China; and National 
Astronomical Observatories, CAS, Beijing, P.R. China; 
yugao@pmo.ac.cn,ygao@nrao.edu}
\altaffiltext{2}{National Radio Astronomy Observatory, 
Socorro, NM 87801, USA; ccarilli@nrao.edu}
\altaffiltext{3}{Department of Physics \& Astronomy, SUNY at Stony Brook,
Stony Brook, NY 11794, USA; philip.solomon@sunysb.edu}
\altaffiltext{4}{National Radio Astronomy Observatory, 
Charlottesville, VA 22903, USA; pvandenb@nrao.edu}

\begin{abstract}

We present here the sensitive HCN(1-0) observations made with the VLA of 
two submillimeter galaxies and two QSOs at high-redshift. HCN emission 
is the signature of dense molecular gas found in GMC cores, the actual 
sites of massive star formation. We have made the first detection of HCN 
in a submillimeter galaxy, SMM J16359+6612.   The HCN emission is seen 
with a signal to noise ratio of 4$\sigma$ and appears to be resolved 
as a double-source of $\approxlt\  2''$ separation. Our new HCN observations, 
combined with previous HCN detections and upper limits, 
show that the FIR/HCN ratios in these high redshift sources lie 
systematically above the FIR/HCN correlation established for nearby 
galaxies by about a factor of 2. Even considering the scatter in the 
data and the presence of upper limits, this is an indication that the 
FIR/HCN ratios for the early Universe molecular emission line galaxies 
(EMGs) deviate from the correlation that fits Galactic giant molecular 
cloud cores, normal spirals, LIRGs, and ULIRGs.  This indicates that the 
star formation rate per solar mass of dense molecular gas is higher in 
the high-$z$ objects than in local galaxies including normal spirals 
LIRGs and ULIRGs.  The limited HCN detections at high-redshift show that 
the HCN/CO ratios for the high-$z$ objects are high and are comparable to 
those of the local ULIRGs rather than those of normal spirals. This 
indicates that EMGs have a high fraction of dense molecular gas compared 
to total molecular gas traced by CO emission. 
\end{abstract}


\section{Introduction}

A molecule whose emission line luminosity  is linearly related to the 
rate of star formation in a galaxy would be useful to the study of star 
formation and the evolution of galaxies since the emission can be used 
to directly and simply relate the star formation rate to physical 
properties of the ISM.  The dust continuum emission in the
far-infrared (FIR),  re-radiated from dust heated 
by UV radiation from young massive stars, is a good indicator of the 
actual star formation rate. A molecular line tracer 
of star formation can be imaged at sub-arcsecond resolution 
and analyzed for the morphology and kinematics of the 
star-forming gas, whereas measurements of the FIR 
lack comparable spatial resolution. 
 
Emission in the rotational lines of carbon 
monoxide is the cardinal indicator of the 
presence of molecular gas in galaxies, seen from the interstellar 
medium of the Milky Way to quasars at redshifts up to $z \sim 6$ 
(see Solomon and Vanden Bout 2005, for a review of the properties 
of the $\sim 40$ galaxies with CO emission detected at $z \geq 1$).  
CO emission is strong from  star-forming giant molecular 
clouds (GMCs) in the Galaxy and external galaxies. And the FIR 
luminosity, a measure of the star formation rate, is directly proportional to 
CO luminosity for GMCs and normal spirals, 
but increases more rapidly than linearly with CO luminosity for 
luminous and ultraluminous infrared galaxies (LIRGs \& ULIRGs, see 
Sanders and Mirabel 1996).  Gao and Solomon (2004a, hereafter GS04a) 
find a slope of 1.7 over 3 orders of magnitude in luminosity for the 
\lir\  -- \lco\  correlation including all of the CO observations of 
ULIRGs from Solomon \etal (1997).  Thus for extreme starbursts such as 
ULIRGS the star formation rate per solar mass of total molecular gas 
traced by CO, a measure of the star formation efficiency, is much higher 
than for normal spirals. 

In contrast to the FIR--CO correlation, the FIR--HCN correlation is a linear 
relation over three decades in HCN luminosity from normal spirals to 
ULIRGs (GS04a). This is based upon the HCN survey 
of 65 galaxies, including nearly 10 ULIRGs and more than 20 LIRGs 
(Gao \& Solomon 2004b; Solomon, Downes, and Radford 1992).  Indeed, 
Wu \etal (2005) have shown that the FIR--HCN correlation for galaxies
extends down to individual dense cores in the star-forming GMCs in 
the Milky Way, spanning over eight decades in HCN luminosity. The key 
point is that HCN emission traces  the gas actually undergoing star 
formation which is $\sim$ 100 times denser than that traced by CO and 
the star formation rate is linearly proportional to the mass of dense 
gas  with approximately n(\h2 ) $ \approxgt\  3 \times 10^4 $ cm$^{-3}$ 
(Solomon, Downes, \& Radford 1992; GS04a)
The work reported here addresses the question of whether this linear 
FIR--HCN correlation, well-established for galaxies with redshifts 
$z < 0.1$, extends to EMGs at high-$z$. 

Previous to this work, there were four high-$z$\  detections of HCN 
(Solomon \etal 2003; Vanden Bout \etal 2004; Carilli \etal 2005; 
Wagg \etal 2005) together with upper limits to five others 
(Isaak \etal 2004; Carilli \etal 2005; Greve \etal 2006).
The FIR/HCN ratios for these galaxies are all systematically offset from 
an extrapolation of the FIR--HCN relation for nearby normal spirals, LIRGs, 
and ULIRGs. Because an accurate molecular 
line tracer of star formation  in EMGs is of great interest, 
it is important to know whether the EMGs truly lie above the 
FIR--HCN correlation observed for $z < 0.1$ or not.  
The four high-$z$ \ detections and upper limits on 5 others lie at the 
limits of current telescope capabilities and the error bars can only 
be reduced with great 
expenditure of integration time. In addition, all four HCN detections
at high redshift are quasars where the dust-enshrouded 
AGN contribution to the FIR emission 
could be significant.  We attempted to 
increase the number of high-$z$ \ HCN detections by observing two 
submillimeter galaxies, neither with an embedded AGN, and two more 
quasars with known strong CO emission.

\section{Observations and Results}
 
The galaxies selected for HCN(1-0), rest frequency 
$\nu$=88.63185 GHz, observations were four EMGs  from
the compilation of Solomon and Vanden Bout (2005) that had the 
strongest CO emission following that of those previously detected 
in HCN emission. In addition, their redshifted 
HCN lines fell into an observing band of the Very 
Large Array (VLA)\footnote{The Very Large Array is a facility of the 
National Radio Astronomy Observatory, operated by Associated Universities, 
Inc., under a cooperative agreement with the National Science Foundation.}.  
The observations were carried out in the D configuration of the VLA 
in 2005 October and December for a total of 54 hours.  A summary of 
the sources, observing times and observed frequencies is given in Table 1.
 
The lowest redshift source in our sample, a submillimeter (SMG) galaxy, 
SMM J02396-0134 at $z=1.062$, 
was observed with the VLA 40-50 GHz receiver system in the 50 MHz 
continuum mode, which has two intermediate frequencies (IFs)
of two polarizations each.  Observations were made with two IF settings 
to provide the bandwidth required to cover the expected HCN line, as 
the CO line width is over 700 \kms\  broad.  This provided four 
independent channels of 350 km s$^{-1}$ each, two covering 
the expected HCN line 
and one each above and below the HCN line for a measurement of the continuum. 

The redshifted HCN lines for other three sources fell in 
the VLA 21-25 GHz receiver system.
These sources have CO line widths from 340 to 500 \kms\  (25 to 37 MHz).  
We observed these sources in the VLA 2AC mode with 7 channels of width 
3.125 MHz (40 \kms) in each of the consecutive IF pair, providing 
a total of $\sim$550 \kms\ velocity coverage. 
Standard VLA data reductions were performed in AIPS. And HCN channel
maps were obtained for each target with the AIPS task `imagr'. 
A summary of the observations and results is given in Table 2.  

No significant HCN emission was seen in any of the targets except for 
the weak line at 4$\sigma $ in SMM J16359+6612 (hereafter J16359), 
the other SMG in the sample.  
We obtained the velocity-integrated moment-zero map 
for each target by integrating over the channel maps covering the 
expected HCN line width, assumed to be the same
as that of CO, leaving out the edge channels. The SMG J16359 shows
 weak but significant HCN emission at the strongest CO position
(Fig. 1). The rms noise 
levels from these moment-zero maps were used to calculate the 3$\sigma$
upper limits to the velocity-integrated HCN flux in the three
undetected sources.

\subsection{HCN(1-0) Emission in SMM J16359+6612}

Figure 1 presents the HCN image of J16359.  The SMG J16359 has 
three widely separated lensed components 
detected in CO (Kneib \etal 2005).  We see a weak HCN signal at 
3$\sigma$ in the strongest of these CO lens components.  
To increase the signal-to-noise ratio, we extracted three HCN images 
of size $\sim 0.54'$ centered at the centroids (crosses) of each of 
the three CO lensed image components in Fig. 1. The inset image is 
a weighted average of these three HCN images.  The (weighted) average 
HCN image shows  a 4$\sigma$ signal at the strongest CO position.  
This HCN source is barely 
resolved as a double source with a separation of $\approxlt\  2''$.  
 There is also a 3$\sigma$ HCN signal $16''$ NE at the location of the 
second strongest CO lens component.
 
The reality of this tentative HCN detection in the component-averaged 
image of J16359 is further supported by the following features evident 
in the CO and HCN morphologies/kinematics:

1. The  HCN emission maxima ($\ge 3\sigma$) are exactly
at the centroids of the two strongest lensed CO components. 
And the strongest CO centroid is located in the middle of
the double HCN components separated by $\approxlt\  2''$. 

2.  The location, orientation,
and separation of the HCN double-source are also consistent
with those of inferred CO kinematics/morphology though the 
CO emission is spatially unresolved. They are also possibly associated with 
the optical features both spatially and kinematically (Kneib \etal 2005).

3. Further indirect evidence for a merger origin of J16359 
has been summarized by Wei\ss~ \etal (2005). Although high
resolution CO and high sensitivity HCN imaging is required 
to directly reveal the putative 
merging morphology (or extended $\sim$\ 2 kpc edge-on dense gas ring/disk), 
the stacked HCN image provides the first direct evidence of a double-source 
in dense star-forming gas. We note that a double-source distribution in 
a lens image needs not imply the same distribution in 
the original object.


\subsection{FIR--HCN Correlation in High-redshift Galaxies}

Combining all HCN observations 
at high-redshift,  including upper limits, yields a total of five 
HCN detections and eight upper limits, shown in Table 3.
This sample contains four SMGs. We note that all of the HCN luminosities 
in Table 3 are for the (1-0) line.

HCN luminosities are calculated
(Solomon \etal 2003)  from 
$L'_{\rm HCN} = 4.1\times10^{3} S_{\rm HCN}\Delta v (1+z)^{-1}D_{\rm L}^2$, 
in K \kms pc$^{2}$, where $S_{\rm HCN}\Delta v$ is the velocity 
integrated HCN line flux in Jy \kms , and $D_{\rm L}$ is the luminosity 
distance in Mpc.  The HCN, CO, and FIR luminosities for our sample, 
are given in Table 3, together with the lens magnification that was 
assumed in correcting apparent to intrinsic luminosities.

In Figures 2-4 we show plots of
FIR vs. HCN luminosity, the FIR/CO  ratio vs. 
the HCN/CO luminosity ratio, and FIR luminosity vs. the HCN/CO  ratio, 
respectively, for the 
13 high-$z$ \ galaxies of Table 3 together with those of a sample of 65 
galaxies from the local HCN survey (GS04a).

\section{Discussion}

The five measurements and eight upper-limits to the HCN luminosities 
in the high-$z$ \ EMGs shown in Figures 2 and 3, 
have systematically somewhat larger FIR luminosities than predicted by 
an extrapolation of the correlation that applies to local
normal spirals, LIRGs, and ULIRGs. This hints at a slight deviation from 
linearity above a FIR luminosity  \lfir\ $> 2.5 \times 10^{12}\  \ls$\ 
that is similar to the much larger non--linearity of the FIR--CO 
correlation particularly for LIRGs and ULIRGs with  
\lfir\ $ > 1 \times 10^{11}$  \ls .

The local sample has an average ratio 
\lfir /\lhcn\ $=$ 750 \ls /  K \kms pc$^{2}$\ for all 65 galaxies over 
a range of luminosity from 2 $\times 10^9$ to 2 $\times 10^{12}$ \ls.   
The ULIRGs in the sample which have FIR luminosities  
\lfir\ $ > 7 \times 10^{11}$ \ls\  (\lir $ > 9 \times 10^{11}$\ls\ ),  
have an average ratio of 880 with a 1$\sigma$ dispersion of 
$(+ 330, -250)$ \ls /  K \kms pc$^{2}$\ . 
These are the closest analogs to the EMGs.
Two of the  high-$z$ \ HCN detections (Cloverleaf and SMM J16359) lead 
to \lfir /\lhcn values within the range expected from nearby galaxies, 
similar to the ratio in the best known ULIRG Arp~220 with 
\lfir /\lhcn\ $=$ 1340 \ls /  K \kms pc$^{2}$\ . The three other detections 
have ratios about twice that of Arp 220.  Only three of the 65 local 
galaxies have ratios in this range (2200 $-$ 2600).
Four measurements yield high  upper limits to \lfir /\lhcn, outside 
the distribution expected from local galaxies including ULIRGs. 
There are also three weak upper limits that are  not very useful. 
Combining the detections and upper limits, the star formation rate per 
solar mass of dense gas, measured by \lfir /\lhcn, is higher in EMGs than 
in the local Universe, including ULIRGs, by a factor of about 2.0 -- 2.5. 

It should be noted that with the exception of four SMGs, including 
the  HCN detection reported here, all other EMGs in our sample
host known AGN. Were the FIR 
luminosities of these EMGs to be corrected by the amount contributed 
by their AGN, the points plotted would, in principle, come into better 
agreement with the low-$z$ \ correlation.  However, for the three quasars 
where the corrections have been calculated (F10214+4724, Downes
\etal 2004; Cloverleaf, Wei\ss\ \etal 2003; APM~08279+5255, 
(Wei\ss\ \etal 2005, 2007) the corrections are significant only 
for APM~08279, a highly luminous IR source with an unususal hot dust 
component (200K) connected with the QSO.  Note, here we have used  
the rest frame FIR for all galaxies, local and high-$z$,  instead of the 
IR luminosity that was used in GS04a for the local HCN sample. 


Table 3 and Fig. 4 show that the ratio \lhcn /\lco\ for the EMGs is similar 
to that found in the local ULIRG population with the highest 
\lhcn /\lco\ $= 0.27$ and an average of 0.19 (for 5 detections) compared 
with  a maximum of 0.26 and an average of 0.13 for local ULIRGs. If we 
leave out APM~08279  the highest EMG value is 0.19 and the average is 0.13.  
Normal spiral galaxies have an average ratio of 0.03. Thus EMGs and ULIRGs  
have in common a high fraction of dense molecular gas compared to total 
molecular gas. All of the EMGs meet the luminous starburst criteria (Fig. 4) 
found by GS04a. 
Kohno (2005) claimed that nearby AGN tend to have exceptionally high 
HCN/CO ratios, presumably due to the X-ray dominated regions 
near AGN. But we did not find any similar cases
in the AGN-dominated EMG sample at high-$z$ as their HCN/CO ratios are  
comparable to that of local ULIRGs (Fig. 4). 
HCN observations of the local host galaxies of infrared-excess 
Palomar-Green QSOs also show that there is no evidence 
that the global HCN emission is enhanced relative to CO in galaxies 
hosting luminous AGN (Evans \etal 2006).

In conclusion we find that the EMGs have large quantities of dense 
molecular gas with HCN luminosities on average higher than local ULIRGs. 
They also have a high dense gas mass fraction, similar to ULIRGs.  But, 
on average, they have a higher star formation rate per unit dense 
molecular gas than ULIRGs or local normal large spirals. 

Significant  progress requires the power of the Atacama Large 
Millimeter Array (ALMA), which will dramatically increase both the 
sample size and the quality of the measurements,  to obtain rotational ladders,
refine density estimates, and
 test other molecules such as HCO$^{+}$, HNC, CS, etc., for 
  measuring the physical 
properties of dense star-forming gas at high-$z$.

\acknowledgements

Yu Gao acknowledges the NRAO, particularly Chris Carilli, for 
hospitality shown. Support for this project came from NSF of 
China and Chinse Academy of Sci. (YG), and 
the Max-Planck Society (CC).  CC and YG thank the Max-Planck Society and 
the Humboldt-Stiftung for partial support through the 
Max-Planck-Forschungspreis 2005.

\begin{deluxetable}{lrrrrr}
\tablenum{1}
\tablewidth{6in}
\tablecaption{Sources Observed \label{tbl-1}}
\tablehead{
\colhead{Source}        &     \colhead{$z$} &  \colhead{R. A. } &
\colhead{Dec. } & \colhead{$t_{\rm obs}$} & \colhead{$\nu_{\rm{obs}}$} \\
 & &  \colhead{(J2000)} & \colhead{(J2000)} &\colhead{(Hours)} &

\colhead{(GHz)}
}
  
\startdata
 
 SMM J02396-0134 &  1.062 &  02:39:56.59 & $-$01:34:26.6  & 21 &  43.0\nl
 SMM J04135+1027 &  2.846 &  04:13:27.50 & +10:27:40.3  &  3 &  23.0\nl
 RX J0911.4+0551 &  2.796 &  09:11:27.50 & +05:50:52.0  & 14 &  23.3\nl
 SMM J16359+6612 &  2.517 &  16:35:44.15 & +66:12:24.0  & 16 &  25.2\nl
 
\enddata
\end{deluxetable}

\begin{deluxetable}{lrrrr}
\tablenum{2}
\tablewidth{6in}
\tablecaption{Observational Parameters and Results \label{tbl-1}}
\tablehead{
\colhead{Source}  &            
\colhead{Beam}       &       \colhead{Noise\tablenotemark{ (1)}}  &
\colhead{Ch. Width}          &
\colhead{$S_{\rm HCN} \Delta v$\tablenotemark{ (2)}} \\
\colhead{ }                  &
\colhead{FWHM ($''$)}       &       \colhead{($\mu$Jy/beam)}  &
\colhead{(MHz, \kms)}          &
\colhead{(mJy \kms)}}
  
\startdata
 
 SMM J02396-0134   & 2.0$\times 1.7$ & 105 & 50, 348.7 & $<$73 \nl
  &    4.0$\times$4.0 & 75 &  & $<$92 \nl
 SMM J04135+10277  &  3.5$\times 3.1$ & 76 & 3.1, 40.6 & $<$63 \nl
 RX J0911.4+0551  &  3.7$\times 3.3$ & 62 & 3.1, 40.1 & $<$29 \nl
 SMM J16359+6612  &  3.0$\times 2.5$ & 96 & 3.1, 37.2 & 33 \nl
 
\tablenotetext{1}{The root-mean-square noise in the channel maps.}
\tablenotetext{2}{Upper limits (3$\sigma$ in moment zero maps) are 
3 times integrations of the noise over the frequency channels where 
the HCN line was expected from observed CO lines.}
\enddata
\end{deluxetable}

\begin{deluxetable}{lrrrrrrr}
\tablenum{3}
\tablewidth{6.8in}
\tablecaption{Intrinsic Properties from HCN (1-0) Detections and Upper Limits\tablenotemark{*}}
 
\tablehead{
\colhead{Source}    &      \colhead{\lfir/\lhcn} &      \colhead{\lfir} &
\colhead{\lhcn}       &       \colhead{\lco}  &
\colhead{\lhcn/\lco}          &
\colhead{Lens} & \colhead{Refs.\tablenotemark{\ddagger}} \\
\colhead{ }         &  \colhead{(\ls/$L_l$\tablenotemark{\dag} )} &  \colhead{($10^{12}$\ \ls)} &
\multicolumn{2}{c}{($10^9\ L_l$\tablenotemark{\dag} )}  & \colhead{}  &
\colhead{Mag.}}
  
\startdata

 VCV J1409+5628  & 2570 & 17  & 6.5  &  74   &   0.09  &    1 & 2, 3, 4 \nl
 APM 08279+5255  & 1000 & 0.25  & 0.25  &  0.92   &   0.27  &  80 & 5, 6, 7 \nl
  H1413 (Cloverleaf)  & 1690 & 5.0  & 3.0  &  37   &   0.08  &  11 & 4 \nl
  IRAS F10214+4724  & 2770 & 3.4  & 1.2  &  6.5    &   0.18  &    17 & 4, 8 \nl
  J16359+6612(B)  & 1430 & 0.93  &  0.6 &  3.7    &   0.18  &    22 & 1, 4 \nl
  BR 1202-0725  & $>$1405 & 55  &  $<$39 &  93   &  $<$0.42  &    1 & 9, 10 \nl
  SMM J04135+1027  & $>$800 & 22  & $<$28 & 159   & $<$0.18  &   1.3 & 1, 4 \nl
  SMM J02399-0136 & $>$609 & 28 & $<$46 & 112 & $<$0.41 & 2.5 & 4, 11 \nl
  SDSS J1148+5251  & $>$2200 & 20 &  $<$9.3 &   25   &  $<$0.36  &    1 & 2, 3, 4 \nl
  SMM J02396-0134  & $>$1625 & 6.1  &  $<$3.7 &  19  &  $<$0.20  &   2.5 & 1, 4 \nl
  SMM J14011+0252  & $>$2500 & 0.7-3.7  & $<$0.3-1.5   &  4-18  &  $<$0.08  &  25-5 & 3, 4 \nl
  MG0751+2716  & $>$2900 & 2.7   & $<$0.9   & 9.3   &  $<$0.10    &   17 & 3, 4 \nl
  RX J0911+0551  & $>$3280 & 2.1  &  $<$0.6  & 4.8    & $<$0.13  &   22 & 1, 4 \nl  

\tablenotetext{*}{Calculated using $H_{\rm 0} = 75$ \kms\,Mpc$^{-1}$, 
$\Omega_\Lambda=0.7$, $\Omega_m=0.3$, and lens magnifications listed.}

\tablenotetext{\dag}{$L_l = $\ K km s$^{-1}$ pc$^{2}$.}

\tablenotetext{\ddagger}{References -- (1) This paper; (2) Beelen et al. (2006); (3) Carilli et al. (2005); (4) Solomon and Vanden Bout (2005); (5) We adopt the  Wei\ss\ et al. (2007) detailed two component model of the CO and HCN excitation and use the predicted HCN(1-0) and FIR luminosity from the collisionally excited gas in the dense cold (65K) star forming component. The observed HCN(5-4) line is produced by the unusually hot dust at T $=$ 200K; (6) Wagg et al. (2005); (7) Egami et al. (2000); (8) Downes and Solomon (2004); (9) 
Riechers et al. (2006); (10) Isaak et al. (2004);  (11) Greve et al. (2006).}

\enddata
\end{deluxetable}


\begin{figure*}
\epsscale{0.68}
\vskip -0.5in
\plotone{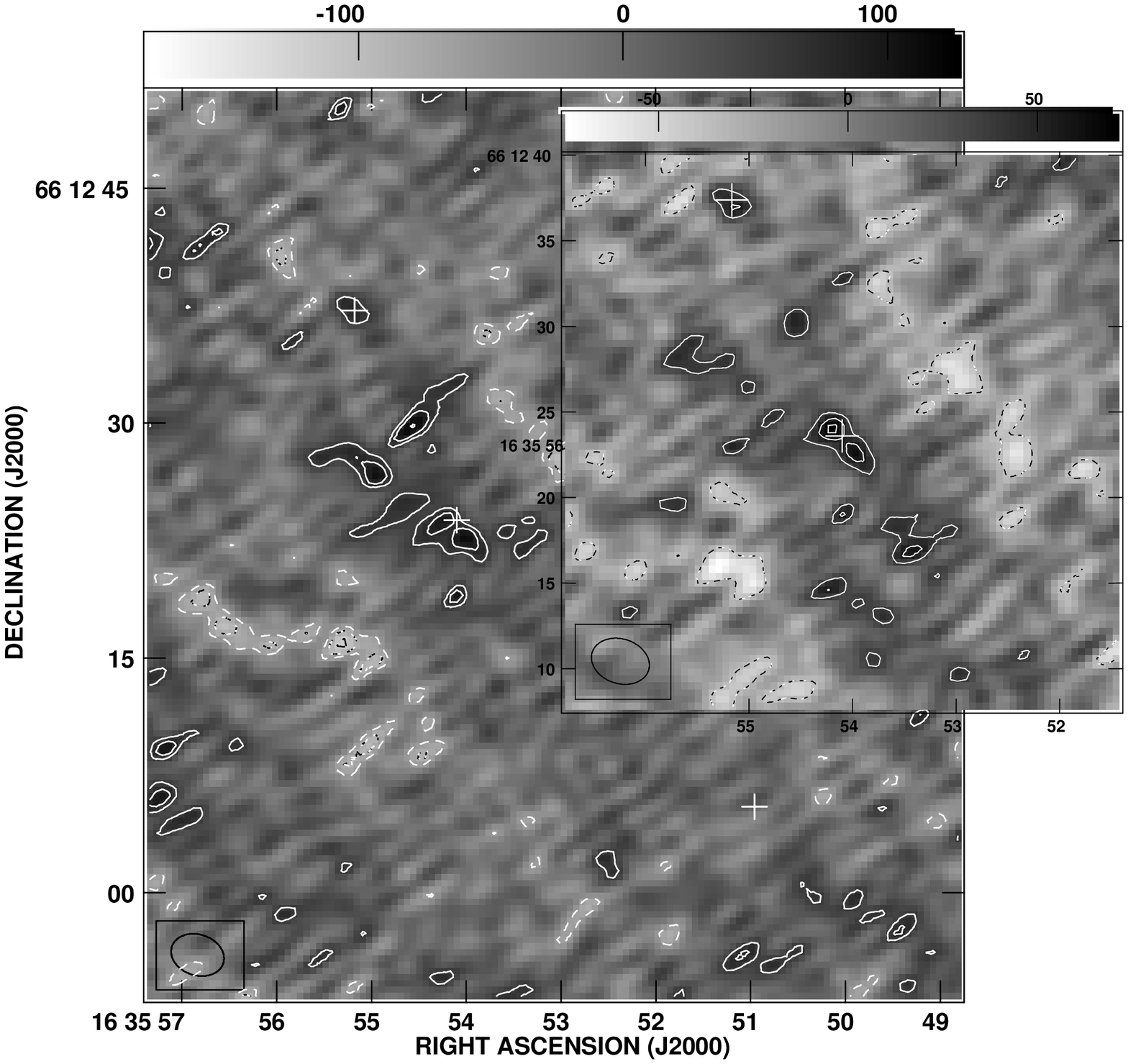}
\figcaption{The larger image shows HCN emission from SMM J16359+6612, 
visible at the position of the CO component B (central cross:
RA=16$^h$35$^m$54.$^s$1, DEC=66$^{\circ}$12$'$23.$''$8). Contours plotted are 
-2.8, -2, 2, 2.8, and 4$\times 12.7$ mJy/beam \kms. 
The inset stacked image (see text) shows a $\sim 4 \sigma$ 
detection at the CO 
component B. Contours plotted are -2, 2, 3, and 4$\times 8$mJy/beam \kms. 
Note the CO component A (NE cross) also shows possible 
HCN ($\sim 3 \sigma$) detection.  
\label{fig1}}
\end{figure*}
 
\newpage
 
\begin{figure*}
\epsscale{0.8}
\vskip -0.5in
\plotone{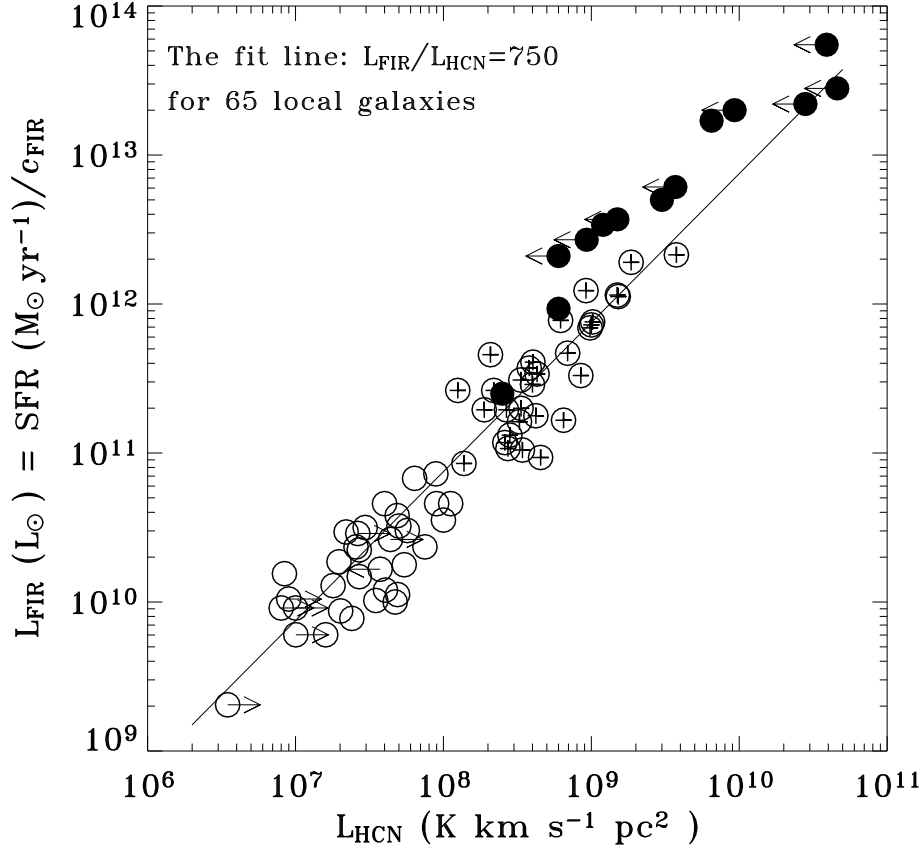}
\figcaption{The correlation between HCN and FIR
luminosities in 13 EMGs and 65 local galaxies (GS04a, note FIR
is used here). The high-$z$ EMGs are indicated in solid
circles, local LIRGs and ULIRGs in crosses, 
and normal spirals in open circle. Limits in HCN luminosities
are indicated with arrows. The line is the fit to the entire 65 local
galaxies with a slope fixed at unity (same as the formal slope of the 
fit: 0.99). Note that the local LIRGs and 
ULIRGs fit this line but the EMGs lie above the line by a factor 
of 2--2.5 indicating a higher star formation rate per solar mass 
of {\bf dense} molecular gas.
\label{fig2}}
\end{figure*}
 
\newpage
 
\begin{figure*}
\epsscale{0.8}
\vskip -0.5in
\plotone{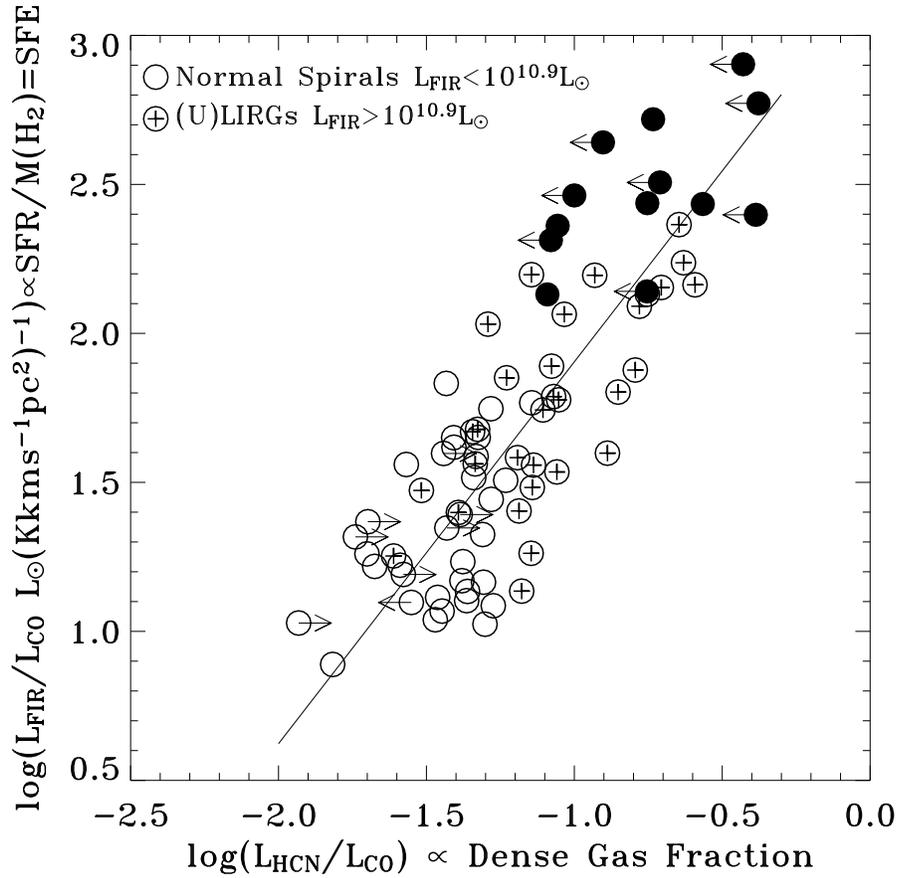}
\figcaption{Correlation between $L_{\rm HCN}/L_{\rm CO}$ and $L_{\rm
FIR}/L_{\rm CO}$ revealing the physical relationship between the HCN 
and FIR since both luminosities are normalized by \lco,  removing the
dependence on distance and galaxy size. The line is the best fit for the local
sample of GS04a. The high-$z$ EMGs (solid circles) show some FIR/CO excess.
\label{fig3}}
\end{figure*}
 
\newpage
 
\begin{figure*}
\epsscale{0.8}
\vskip -0.5in
\plotone{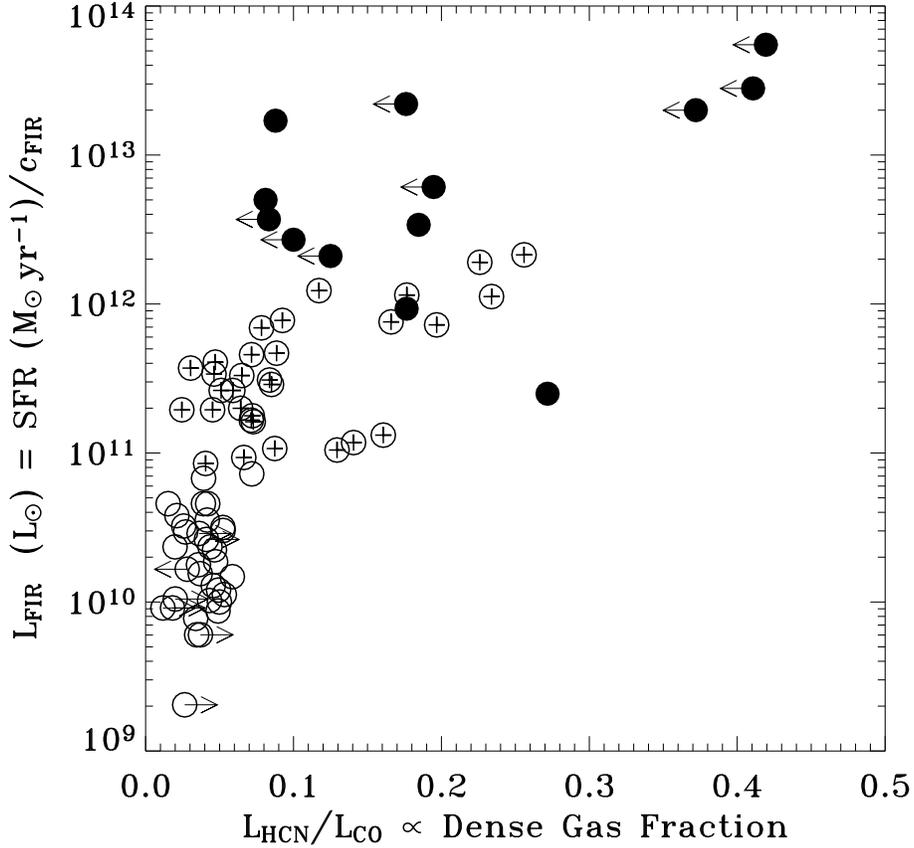}
\figcaption{\lfir \  vs. $L_{\rm HCN}/L_{\rm CO}$. 
All galaxies 
with a high dense molecular gas fraction of 
$L_{\rm HCN}/L_{\rm CO} > 0.06$, whether local or 
high-$z$, are IR-luminous ($L_{\rm FIR}>10^{11}$ L$_{\sun}$).
The high-$z$ EMGs are indicated in solid
circles, local LIRGs and ULIRGs in crosses, 
and normal spirals in open circles. The high-$z$ EMGs show a high ratio of dense to total molecular gas similar to ULIRGs but even higher FIR luminosity.  
\label{fig4}}
\end{figure*}

\end{document}